# Cascade solutions of the Lorenz system

Zeling Chen, Hong Zhao*

*Department of Physics, Xiamen University, Xiamen 361005, China*

*\*zhaoh@xmu.edu.cn*

The Lorenz system is a milestone model of nonlinear dynamic systems. However, we report in this Letter that important information of the global solutions in the parameter space may still miss: there is a series of cascade solutions in certain regions of the parameter space that has not been found before. We denote them as $S_i$, where i=1,2,3,…, can tend to be infinite. The solution $S_1$ is the fundamental solution having been studied intensively, the second one has also been reported and studied partially before. We conduct a detailed study on other solutions to reveal their special self-similar features, which enable us to find that solutions of i=1,3,5,…, and solutions with i=2,4,6,…, have different symmetry, but in each series all of the solutions are qualitatively similar after properly rescaled. The phase diagram of cascade solutions is shown in the parameter space. Odd and even solutions may coexist in certain parameter regions, in which case their basins of attraction show fractal geometry. With high-order solutions, the Lorenz system is shown analog to an excitable system. An ordinary differential equation existing a series of infinite cascade solutions has never been reported before.

The Lorenz system is one of the canonical models in studying nonlinear dynamics behaviors[1]. After the presentation of this model, numerous nonlinear systems were introduced to study the chaotic properties[2-6]. Besides the theoretical importance, this model has a number of important applications in many fields such as mechanics [7-8], optics[9], chemistry[10-11], economics[12-14].

The Lorenz system is given by three ordinary differential equations, $\dot{x} = \sigma(y-x)$, $\dot{y} = x(r-z) - y$, $\dot{z} = xy - bz$, where $\sigma$, $r$, and $b$ are parameters. The essential feature is that by varying the parameters, the attractors of this model vary in form. In particular, for certain values of the parameters the model exhibits periodic motion, while for other values it exhibits chaotic behavior. From periodic to chaotic motion it follows the route of period-doubling bifurcation, and then develops to full chaotic states with symmetry-break solutions, which thus can show more complex chaotic motions that cannot exist in other canonical models such as the one-dimensional Logistic map and the two-dimensional Hénon map [15].

This system has three fixed points, $(x,y,z) = (0,0,0)$, $(x_0, y_0, z_0)$ and $(-x_0, -y_0, z_0)$, where $x_0 = y_0 = \sqrt{bz_0}$ and $z_0 = r - 1$. It is found that after these fixed points lose their stability, trajectories pass around the two points, $(x_0, y_0, z_0)$ and $(-x_0, -y_0, z_0)$, of attraction from one to another in a way of periodic or in a chaotic fashion. Extensive previous researches have focused on this kind of solution which dominate the major region of the parameter space. Meanwhile, several studies have found hints that this system exists another solution in certain specific parameter regions. Sparrow[16] first observes a new solution (at $\sigma = 10$, $r = 120$, and $b = 0.1$) having a different topology from the major solution studied before, i.e., instead of switching from one to another attraction of unstable fixed points the trajectory moves around just one of the fixed points. He recognizes the importance of the Lorenz system in parameter region of small b and guesses that this region should have most complicated behavior ever observed. Later Alfsen and Frøyland[17] study this kind of solution in certain detail, and find that the periodic trajectory may develops to chaos by also the period-doubling bifurcation route. Some researchers[18] have studied the case of negative b and observed similar solutions.

In this letter, we report that indeed the Lorenz system exists a set of cascade solutions in the parameter space with quite small b. We denote them by $S_i$ with i=1,2,3,…. The fundamental solution that has been studied extensively is the $S_1$ solution, the solution found by Sparrow et. al is $S_2$. Solutions in the series i=1,3,5,…, and solutions in the series i=2,4,6,…, have the self-similarity, respectively. That is, by a proper rescaling, the bifurcation diagrams of the solutions in each series appear identical with each other. In the axis of evolution time, however, trajectories of different solutions have difference. A trajectory may stay a much long time when it is close to the z-axis in a high-order solution than in a low-order solution. As a result, the Lorenz system appears as an excitable system. Cascade solutions have been reported in optical feedback systems governed by differential delay equations, where odd harmonic solutions are excited continuously by strange and strange nonlinearity [19-21]. The period of trajectories in harmonic solutions decreases as $T_F/(2n+1)$, where $T_F$ is the period of the fundamental solution, $n$ is the order of the harmonic. In our case of the Lorenz system, however, the period of high-order solutions increases with the increase of the order. To the best of our knowledge, cascade solutions have not been reported in any ordinary differential equations before.

We employ the bifurcation diagram to show the global behavior of the system. To obtain the Poincare section map, however, we adopt a technique slightly different from the conventional way. We record the section points that a trajectory crosses the section of z=r-1 from down to up, instead of from up to down as adopted usually. The reason is that in the case with very small b all of the trajectories from up to down contract to the nearby of the z-axis with coordinates of $|x|\sim|y| < 10^{-7}$ usually, leading to the section points indistinguishable. On the contrary, section points obtained by the modified way have high distinguishability since the trajectories are far from the z-axis. We wonder whether this is the reason that the cascade solutions have not been found systematically before.

Figure 1 shows the bifurcation diagram along the parameter b with other parameters fixed at the well-known parameter values of $\sigma$ =10 and $r$ = 28. The vertical axis is for the section points of $x_c$ at $z_c = r - 1$. The horizontal axis is shown with a log scale to display the solutions in the region of small b. To show the coexisting attractors we obtain the bifurcation diagram by first searching for the section points from large b to small b with a proper initial condition, and then perform the search from small b to large b again. In such a way we can obtain the coexisting attractors. The cascade solutions are clearly shown, by which one can infer that the series should be extended to infinite.

The fundamental solution $S_1$ has been extensively studied. It always has two branches (characterized by the same color) in our bifurcation diagram, which implies that a trajectory would wander between the attractions of the two unstable fixed points of $(x_0, y_0, z_0)$ and $(-x_0, -y_0, z_0)$. There are windows with symmetry-break attractors, red-color and blue color branches of the bifurcation diagram, but still, trajectories of each branch are trajectories wandering between the regions of the two fixed pointes.

The $S_2$ represents another solution with different symmetry. In the initial stage of the bifurcation diagram, the up and down branches are symmetry-break attractors. Trajectories on each branch move in the attraction of one of the unstable fixed points, $(x_0, y_0, z_0)$ or $(-x_0, -y_0, z_0)$, respectively. Each branch develops to chaos by the period-doubling bifurcation, and then the two branches combine as one branch after colliding at a crisis parameter point.

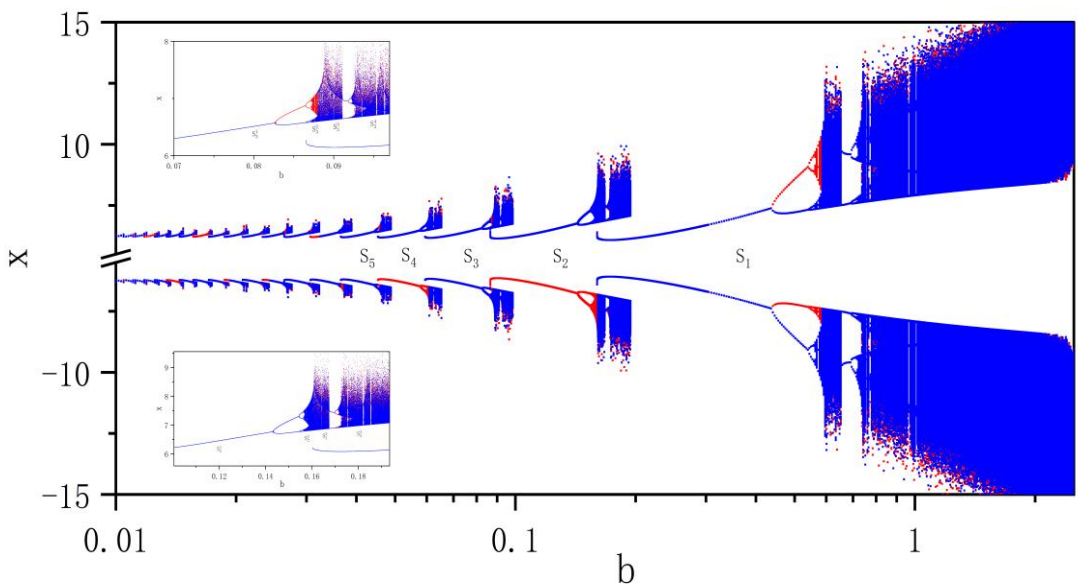

FIG.1 The bifurcation diagram of the cascade solutions along the b axis. The other parameters are fixed at σ =10 and r = 28. The blue and red color points symmetry-break solutions.

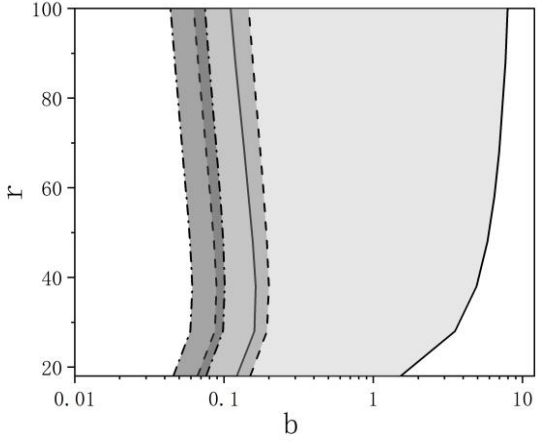

FIG.2 The phase diagram of cascade solutions for the first three solutions in the parameter plane of r-b. The solid, dashed, and dot lines bound the boundaries of S1, S2, and S3, respectively.

A very interesting finding is that the cascade solutions have well defined self-similarity, that is, solutions of $S_1$, $S_3$, $S_5$,…, have the same symmetry, while solutions of $S_2$, $S_4$, $S_6$,…, have the same symmetry, respectively. By properly rescaling, bifurcation diagrams of different solutions in each set may appear identical. As an example, the top insert in the figure shows the enlargement of $S_3$ of its up branch. Compare with $S_1$, we see that there is no qualitative different between the two bifurcation diagrams. The bottom insert in the figure shows the enlargement of $S_2$ of its up branch. By compare it with other even solutions, one can also confirm that they have the similar bifurcation diagrams.

Figure 2 shows the phase diagram of several cascade solutions in the r-b parameter plane. The axis of b is shown by the log scale. It can be seen that the high-order solutions are located in the region with extremely small b.

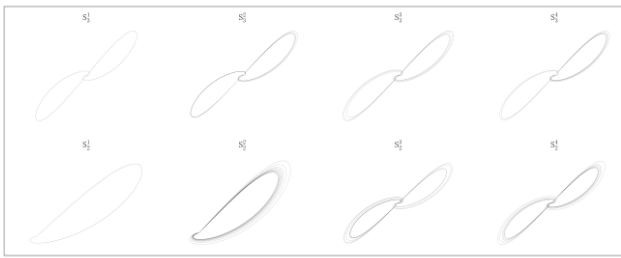

FIG.3 Attractors along with the bifurcation diagrams of S3 (top) and S2 (bottom).

Figure 3 shows several attractors on the x-y plane of $S_2$ and $S_3$ solutions with the parameter positions marked by symbols $S_2^1, S_2^2, S_2^3, S_2^4$ and $S_3^1, S_3^2, S_3^3, S_3^4$ respectively in the bifurcation diagrams (see inserts in Fig. 1). We see that $S_2^1, S_2^2$ show obvious difference in symmetry from $S_3^1, S_3^2$, while $S_2^3, S_2^4$ are qualitatively similar to $S_3^3, S_3^4$. Detail investigations confirm that after the two branches of $S_2$ in the attractions of $(x_0, y_0, z_0)$ and $(-x_0, -y_0, z_0)$ crash at the origin, the structure of the bifurcation diagram of $S_2$ becomes similar to that of $S_1$ after its symmetry-break branches (red and blue color branches) crashes. In addition, we have checked that the well-known scaling constants, such as the Feigenbaum's constant for the period-doubling bifurcations, agree with the universal value having been found in the fundamental solution as well as many other dynamical systems.

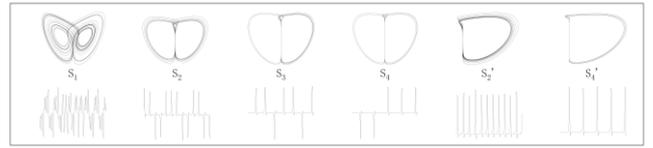

FIG.4 Attractors along the bifurcation diagrams and their time series graph.

Although qualitatively similar, the time evolution of trajectories of higher order cascade solutions has a quantitative difference from those of lower order cascades. The first four plots in the top line of Fig. 4 show attractors in the z-x plane for a fully developed chaotic motion of the first four solutions, and the remained last two plots show chaotic attractors in the z-x plane of the branch around the fixed point $(x_0, y_0, z_0)$ of the $S_2$ and $S_4$ solutions picked up at positions in the symmetry-break regime. We see that with the increase of the cascade order, trajectories constraint toward the region extremely close to the z-axis. The plots at the bottom of the figure show the time evolution of the z variable corresponding to the attractors in the top line. It can be seen that the transition time that a trajectory stays around the z-axis becomes long and long. All of the trajectories spin down around the z-axis slowly when they close to the z-axis while moving from down to up when they leave away from the z-axis. This phenomenon can be understood easily. The eigenvalues of the fixed point (0,0,0) are $\left[-b, \sqrt{ar + \frac{(a-1)^2}{4}} - \frac{1+a}{2}, -\sqrt{ar + \frac{(a-1)^2}{4}} - \frac{1+a}{2}\right]$. It can be easily found that the z-axis is just the stable manifold of the eigenvalue –b. As a result, with the decrease of b, the speed of trajectories spinning down around the z-axis becomes slower and slower. Meanwhile, the speed of contraction of trajectories on the x-y plane at large z keeps almost unchanged, as it is determined by the other two parameters. Therefore, trajectories have a sufficient time to extremely close to the z-axis for high-order solutions since they occur in the parameter regions of quite small b. This is also the reason why we need to trace a trajectory to find the Poincare section points when it passes the section from down to up. In such a way we can catch the section points with large values of x and y. Otherwise, if we trace the trajectory from up to down, the section points we obtained should have quite small values. For example, in the case of the third solution, the section points (xc, yc) obtained in a later way should be within an amplitude of $|xc|\sim|yc|<10^{-5}$.

This phenomenon means that the trajectories of high order solutions, particularly the symmetry-break branches of even order solutions, indeed analog the typical trajectories of excitable systems[22]. In the standard excitable system, trajectories fluctuate in a small region for a long time and then are randomly excited to a large range and then gradually return to the original region. Instead, here trajectories stay in the nearby of the z-axis with x, y variable keep almost zero values for a long time and then being randomly excited to a large range and then gradually return to the nearby of the z-axis again. Therefore, the Lorenz system with high order cascade solutions can be also applied as an excitable medium model.

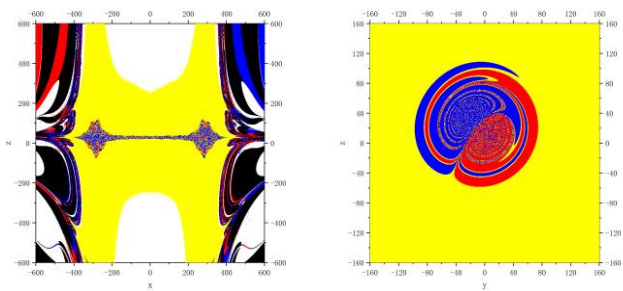

FIG.5 Attraction basins in the plane of z-x at y=1 and z-y plane at x=295.

From Fig. 1 we see that there are many parameter regions in which attractors of different cascade solutions coexist. In this case, their attraction basins have a fractal structure. In Fig.5 we show such an example at the parameter value of b=0.09. Fig.5 (a) shows the x-z plane at the section of y=1.

The blue and red colors represent the attraction basins of the two symmetry-break attractors of $S_2$ around the two unstable fixed points. The yellow color represents the attraction basin of $S_3$. The black color represents the attraction basin of divergent trajectories. The white color represents the attraction basin of trajectories that still stay nearby of the z-axis after an evolution time of $t = 10^6$; it is a reasonable guess that these trajectories will finally drop on one of the coexisting attractors. In Fig.5 (b) we show another section of the phase plane, i.e., the y-z plane at x=295. It is a vertical section at the position marked in Fig. 5(a) by the dashed line. We see that the basins have a complex spiral fractal structure. These sections indicate that the attraction basin of $S_3$ usually lies in the region far from the origin, while the attraction basins of the symmetry-break attractors of $S_2$ mainly lie in the region close to the origin. In the transitional regions, the basins show fractal structures.

In conclusion, we find that the Lorenz system exists cascade solutions in the parameter plane by carefully searching for the parameter region of small b. These solutions show a specific self-similar property: the bifurcation diagrams with odd orders and the bifurcation diagrams with even order are qualitatively identical, respectively. Excepting a proper scaling transformation, the topology structure is accurately the same. The difference between the two alternative solutions are lies in their early region of the bifurcation diagrams, where the odd numbered solutions, including the one has extensively studied before, show attractors globally wander around the two unstable fixed points, while the even numbered solutions show coexisting attractors respectively lie in the regions that the two fixed points located. After their crisis parameter points, trajectories of either solution become global ones, and afterward, the bifurcation diagrams of two adjusted solutions becomes also identical qualitatively.

---